\documentclass[aps,pre]{revtex4}%,groupedaddress,showkeys]{revtex4-1}
\usepackage{graphicx}
\usepackage{subfig}
\usepackage{amsmath}
\usepackage{array}
\begin{document}

\captionsetup[subfloat]{font=normalsize,position=top}

\title{Fluctuations in the Time Variable and Dynamical Heterogeneity
  in Glass-Forming Systems} 

\author{Karina E. Avila$^{1,2}$} 
\author{Horacio E. Castillo$^{1,}$}
\email[]{castillh@ohio.edu}
\affiliation{$^{1}$Department of Physics and Astronomy and Nanoscale and Quantum Phenomena Institute, Ohio
  University, Athens, Ohio, 45701, USA\\ 
$^{2}$Max-Planck-Institut f\"ur Dynamik und Selbstorganisation, Am
  Fassberg 17, D-37077 G\"ottingen, Germany} 

\author{Azita Parsaeian}
\affiliation{Materials Research Center, Northwestern University,
  Evanston, Illinois 60208-3108, USA\\}% 

\date{\today}

\begin{abstract}
We test a hypothesis for the origin of dynamical heterogeneity in
slowly relaxing systems, namely that it emerges from soft (Goldstone)
modes associated with a broken continuous symmetry under time
reparametrizations. We do this by constructing coarse grained
observables and decomposing the fluctuations of these observables into
transverse components, which are associated with the postulated
time-fluctuation soft modes, and a longitudinal component, which
represents the rest of the fluctuations. Our test is performed on data
obtained in simulations of four models of structural glasses. As the
hypothesis predicts, we find that the time reparametrization
fluctuations become increasingly dominant as temperature is lowered
and timescales are increased. More specifically, the ratio between the
strengths of the transverse fluctuations and the longitudinal
fluctuations grows as a function of the dynamical susceptibility,
$\chi_4$, which represents the strength of the dynamical
heterogeneity; and the correlation volumes for the transverse
fluctuations are approximately proportional to those for the dynamical
heterogeneity, while the correlation volumes for the longitudinal
fluctuations remain small and approximately constant.
\end{abstract}

%, 05.40.-a}
%
% From 2010 PACS (http://www.aip.org/pacs/)
% 
% 64.70.Q- Theory and modeling of the glass transition 
% 61.20.Lc Time-dependent properties; relaxation (for glass
% transitions, see 64.70.P-) 
% 61.43.Fs Glasses
% 05.40.-a Fluctuation phenomena, random processes, noise, and
%          Brownian motion (for fluctuations in superconductivity, see
%          74.40.-n; for statistical theory and fluctuations in nuclear
%          reactions, see 24.60.-k; for fluctuations in plasma, see
%          52.25.Gj; 
%          for nonlinear dynamics and chaos, see 05.45.-a) 
% 

\maketitle

\section{Introduction}

The rapidly increasing relaxation timescales, the presence of
non-exponential relaxation, as well as the violation of
Stokes-Einstein relations between viscosity and diffusivity are some
features observed close to the glass transition in glassy
systems~\cite{Debenedetti2001, Ediger2000, Sillescu}. The appearance
of these features suggests that relaxation dynamics is heterogeneous,
i.e. that it is faster in some regions and slower in
others~\cite{Ediger2000, Sillescu, Berthier2011}. Direct microscopic
evidence for this behavior has been found both in
simulations~\cite{Berthier2011, Lacevic} and in
experiments~\cite{Berthier2011, VidalRussell1998, Walther1998,
  Russell2000, Kegel2000, Weeks2000, Weeks2002, Courtland2003,
  Cipelleti, Keys2007, Abate2007, Duri2009}. The understanding of
dynamical heterogeneity is believed to be crucial to explain anomalous
behavior of materials near the glass transition, and even possibly to
explain the very presence of the glass transition
itself~\cite{Ediger2000}. Despite many efforts trying to address the
origin of dynamical heterogeneity, this question still remains
open~\cite{Berthier2011, Toninelli, Garrahan, Lubchenko,
  Castillo2003}.

In recent years, different tools have been used to probe dynamical
heterogeneity. One such tool is the dynamical susceptibility,
$\chi_4$, which depends both on the strength of the local fluctuations
and the spatial extent of their correlations. The peak value of
$\chi_4$ has been observed to grow while approaching the glass
transition. However, this tool by itself cannot tell us much about the
origin of dynamical heterogeneity and it may be desirable to
supplement it with other ways of probing the fluctuations in the
system.

 In this work, we test the predictions of a theoretical framework that
 aims to describe the slow part of the fluctuations in the relaxation
 dynamics~\cite{Castillo2003, Chamon2002, Castillo2002, Chamon2004,
   Chamon2007}. This framework is based on the hypothesis that in
 glassy systems the long time dynamics is invariant under
 reparametrizations of the time variable~\cite{Castillo2003,
   Chamon2002, Castillo2002, Chamon2004, Chamon2007, Castillo2008,
   Mavimbela2011, Franz2010, Franz2011}, but 
 this invariance is broken, giving rise to Goldstone modes that
 manifest themselves with the emerge of heterogeneous
 dynamics~\cite{Castillo2003, Chamon2002, Castillo2002, Chamon2004,
   Chamon2007, Castillo2007, Parsaeian2008, Parsaeian2009,
   Parsaeian2008a, Avila, Mavimbela}. The Goldstone modes correspond to
 fluctuations in the time reparametrization $t \rightarrow
 \phi_{\vec{r}}(t)$, i.e. 
\begin{equation}
C_{\vec{r}}(t,t') \approx C[\phi_{\vec{r}}(t),\phi_{\vec{r}}(t')],
\label{hyp0}
\end{equation}
where $C(t,t')$ is a global two--time correlation function. Some
indirect evidence in favor of the presence of this kind of fluctuation
in structural glasses has been presented in~\cite{Castillo2007,
  Parsaeian2008, Parsaeian2009, Parsaeian2008a}. In this work, we
present results for a more direct test, based on decomposing
fluctuations into a {\em transverse part} satisfying Eq. (1) and a
{\em longitudinal part} containing all other fluctuations. This
procedure allows one to separately quantify the strength and spatial
correlations of both kinds of fluctuations, as a function of
temperature and timescales, for a variety of glass-forming models both
below and above the mode coupling critical temperature $T_c$.  The
same kind of test can also be applied to particle tracking
experimental data from colloidal and granular
systems~\cite{Berthier2011, Kegel2000, Weeks2000, Weeks2002,
  Courtland2003, Cipelleti, Keys2007, Abate2007}, thus allowing to
investigate a possible unified explanation of dynamical heterogeneity
in diverse systems. A summary of the results of an early version of
our analysis was published in Ref.~\cite{Avila}.

This manuscript is organized in the following way. In Sec.~\ref{TRI}
we discuss the hypothesis and define the quantities we use to test
it. In Sec.~\ref{systems} we present the details of the numerical
simulations used to test the hypothesis and present the results of our
analysis. Finally, we discuss our conclusions in
Sec.~\ref{conclusions}.
  
\section{Time reparametrization invariance and fluctuations}\label{TRI} 
We start by discussing in general terms the expected effects of time
reparametrization invariance on the fluctuations. Let us consider the
global two--time correlation function
 \begin{equation}
C(t,t')= \left\langle \frac{1}{N}\sum_{j=1}^N\cos \{\vec{q}\cdot [\vec{r}_j(t)-
  \vec{r}_j(t')]\} \right\rangle,  
\label{global-correlation}
   \end{equation}
where $N$ is the number of particles, $\vec{r}_j(t)$ is the position
of particle $j$ at time $t$, and $\vec{q}$ is the wave-vector that
corresponds to the main peak of the static structure factor. Here
$\langle \cdots \rangle$ denotes an average over thermal fluctuations,
which in our case is approximated by an average over independent
molecular dynamics (MD) runs. In equilibrium, and more generally in
time translationally invariant (TTI) systems, this correlation
function depends only on the difference between the two times
$t-t'$. In the case of aging, the system is no longer TTI and
$C(t,t')$ depends nontrivially on both times $t$ and $t'$.

To analyze the fluctuations in the dynamics we define the local
correlation as~\cite{Castillo2007, Parsaeian2008, Parsaeian2009,
  Parsaeian2008a} 
\begin{equation} 
C_{\vec{r}}(t,t')=\frac{1}{N(B_r)}\sum_{r_j(t')\in
  B_r}\cos\{\vec{q}\cdot[\vec{r}_j(t)-\vec{r}_j(t')]\}, 
\end{equation} 
where the average over particles in Eq. (\ref{global-correlation}) is
now restricted to a region $B_{\vec{r}}$ around point $\vec{r}$, which
contains $N(B_{\vec{r}})$ particles at time
$t'<t$. $C_{\vec{r}}(t,t')$ probes the mobility of the particles in a
region near $\vec{r}$ between times $t'$ and $t$: it is close to zero
for regions where the particle configuration has changed
significantly, and much closer to unity for regions where it has
changed little or not at all.

In this work we will consider a slightly more restrictive version of
Eq. (\ref{hyp0}), namely~\cite{Avila, Bouchaud, Notation} 
\begin{equation}
C_{\vec{r}}(t,t') \approx g[\phi_{\vec{r}}(t)-\phi_{\vec{r}}(t')],
\label{hyp}
\end{equation}
which corresponds to the case when the global correlation has the
form~\cite{Avila, Bouchaud, Notation} 
\begin{equation}
C(t,t') = g[\phi(t)-\phi(t')].
\label{hyp-global}
\end{equation}
In principle, the functions $g(x)$ in Eqs.~(\ref{hyp}) and
(\ref{hyp-global}) could be different. For example, it has been
claimed~\cite{Ediger2000} that stretched-exponential global
relaxation could be the result of combining local exponential
relaxations with different relaxation times, in which case we would
have $g_{\rm global}=A\exp{(-|x|^{\beta})}$ in Eq.~(\ref{hyp-global})
and $g_{\rm local}=A'\exp{(-|x|)}$ in
Eq.~(\ref{hyp})~\cite{Mavimbela}. In this work, for simplicity, we
impose the condition $g_{\rm global}(x)=g_{\rm local}(x)=g(x)$. Both
this restriction and the restriction imposed in Eq.~(\ref{hyp-global})
could in principle make the results appear to be slightly less
consistent with the hypothesis than they would be otherwise. The forms
of the functions $g(x)$ and $\phi(t)$ can be obtained by fitting the
global correlation $C(t,t')$, as we do in the next section. We will
assume, for the moment, that $g(x)$ and $\phi(t)$ are known.

As mentioned before, we refer to the fluctuations described by
Eq. (\ref{hyp}) as {\em transverse fluctuations} and the rest of the
fluctuations as {\em longitudinal fluctuations}~\cite{phi}.  To
visualize these ideas in more detail, we define
\begin{equation}
\Phi_{ab,{\vec{r}}} \equiv g^{-1}(C_{ab,{\vec{r}}}),
\label{hyp2}
\end{equation}
 with $a, b \in \{1,2,3\}$~\cite{Notation}, where $C_{ab,{\vec{r}}}
 \equiv C_{\vec{r}}(t_a,t_b)$. The fluctuating quantity
 $\Phi_{ab,{\vec{r}}}$ is composed of a transverse contribution,
 \begin{equation}
 \Phi_{ab,{\vec{r}}}{}^T=\phi_{\vec{r}}(t_a)-\phi_{\vec{r}}(t_b), 
 \label{transver-eq}
 \end{equation}
 and a longitudinal contribution, $\Phi_{ab,{\vec{r}}}{}^L$, i.e.,
 \begin{equation}
g^{-1}(C_{ab,{\vec{r}}})=\Phi_{ab,{\vec{r}}}=\Phi_{ab,{\vec{r}}}{}^T+\Phi_{ab,{\vec{r}}}{}^L=\phi_{\vec{r}}(t_a)-\phi_{\vec{r}}(t_b)+\Phi_{ab,{\vec{r}}}{}^L. 
\label{hyp3}
\end{equation}
This means that in the absence of longitudinal fluctuations,
Eq. (\ref{hyp}) would be exact.  In order to quantify both kinds of
fluctuations, we define
 \begin{eqnarray} 
{\sigma_{\vec{r}}} & \equiv & \frac{1}{\sqrt{3}}  \left [
g^{-1}(C_{21,{\vec{r}}}) +g^{-1}(C_{32,{\vec{r}}})
-g^{-1}(C_{31,{\vec{r}}})  \right] \nonumber \\ 
& & = \frac{1}{\sqrt{3}} \left( \Phi_{21,{\vec{r}}}{}^{L}
+\Phi_{32,{\vec{r}}}{}^{L}-\Phi_{31,{\vec{r}}}{}^{L} \right), 
\label{sigma}
\end{eqnarray}

\begin{eqnarray}
{\pi}_{1,{\vec{r}}} & \equiv & \frac{1}{\sqrt{2}} \left[
g^{-1}(C_{21,{\vec{r}}})-g^{-1}(C_{32,{\vec{r}}}) \right]
\nonumber \\ 
 & = & \frac{1}{\sqrt{2}}\left[
  (\Phi_{21,{\vec{r}}}{}^{T}+\Phi_{21,{\vec{r}}}{}^{L})-
  (\Phi_{32,{\vec{r}}}{}^{T}+\Phi_{32,{\vec{r}}}{}^{L}) 
  \right],
\label{pi1}
\end{eqnarray}
and

\begin{eqnarray}
{\pi}_{2,{\vec{r}}} & \equiv & \frac{1}{\sqrt{6}}  \left [
g^{-1}(C_{21,{\vec{r}}})+g^{-1}(C_{32,{\vec{r}}}) +2
g^{-1}(C_{31,{\vec{r}}})\right] \nonumber \\ 
 & = & \frac{1}{\sqrt{6}} \left[
  (\Phi_{21,{\vec{r}}}{}^{T}+\Phi_{21,{\vec{r}}}{}^{L})+(\Phi_{32,{\vec{r}}}{}^{T}+\Phi_{32,{\vec{r}}}{}^{L}) 
  +2(\Phi_{31,{\vec{r}}}{}^{T}+\Phi_{31,{\vec{r}}}{}^{L}) \right], 
 \label{pi2}
\end{eqnarray}
with $t_1<t_2<t_3$. As shown above, ${\sigma_{\vec{r}}}$ contains only
longitudinal fluctuations, but $\pi_{1,{\vec{r}}}$ and
$\pi_{2,{\vec{r}}}$ contain both transverse and longitudinal
components.  If Eq.~(\ref{hyp}) was an exact identity, the local
two-time function would verify the following relation:

\begin{equation}
g^{-1}(C_{21,{\vec{r}}})+g^{-1}(C_{32,{\vec{r}}})-g^{-1}(C_{31,{\vec{r}}})=0. 
\label{PhiL=0}
\end{equation}
Then, in the case where no longitudinal fluctuations are present, the
vector $(\sigma_{\vec{r}}, \pi_{1,{\vec{r}}}, \pi_{2,{\vec{r}}})$
would be restricted to be fluctuating in the plane
$\sigma_{\vec{r}}=0$. We expect that as the temperature becomes lower,
the timescales become longer, and the system becomes more glassy,
transverse fluctuations should become progressively more dominant,
according to the hypothesis [Eq.~(\ref{hyp0})] and therefore the
probability distribution $\rho(\sigma_{\vec{r}}, \pi_{1,{\vec{r}}},
\pi_{2,{\vec{r}}})$ should become anisotropic, extending mostly along
the $\sigma_{\vec{r}}=0$ plane and not away from it.

An analogous set of quantities can be defined starting from the global
correlation $C(t,t')$. We can write the analogs of
Eqs. (\ref{transver-eq}) and (\ref{hyp3}), i.e.
\begin{equation}
\Phi_{ab}{}^{T}=\phi(t_a)-\phi(t_b),
\end{equation}
 \begin{equation}
g^{-1}[C(t_a,t_b)]=\Phi_{ab}=\Phi_{ab}{}^T+\Phi_{ab}{}^L=\phi(t_a)-\phi(t_b)+\Phi_{ab}{}^L. 
\label{hyp3-global}
\end{equation}
However, in this case, the meanings of the symbols are
different. $\Phi_{ab}{}^T$ and $\Phi_{ab}{}^L$ do not fluctuate,
because they are computed from the global correlation in the
thermodynamic limit, which is self averaging. If
Eq.~(\ref{hyp-global}) was exact, then $\Phi_{ab}{}^L\equiv
0$. However, in practice Eq.~(\ref{hyp-global}) is only approximate:
in Sec.~{\ref{systems}} we fit the left-hand side by an expression with the form
given in the right-hand side. Thus $\Phi_{ab}{}^T$ represents the part of
$\Phi_{ab}$ that can be represented as a difference
$\phi(t_a)-\phi(t_b)$ according to the fits, and $\Phi_{ab}{}^L$
represents the part of $\Phi_{ab}$ that the fit does not capture, or
in other words a fitting error. Additionally, the global variables
$\sigma$, $\pi_1$ and $\pi_2$ can be defined, by analogy to
Eqs.~(\ref{sigma}), (\ref{pi1}), and (\ref{pi2}), in the following way:

 \begin{equation} 
{\sigma} \equiv  \frac{1}{\sqrt{3}} \left(
\Phi_{21}+\Phi_{32}-\Phi_{31} \right)=\frac{1}{\sqrt{3}} \left(
\Phi_{21}{}^L+\Phi_{32}{}^L-\Phi_{31}{}^L \right), 
\label{sigma-global}
\end{equation}

\begin{equation}
{\pi}_{1} \equiv \frac{1}{\sqrt{2}}\left( \Phi_{21}-\Phi_{32} \right),
\label{pi1-global}
\end{equation}
and

\begin{equation}
{\pi}_{2} \equiv \frac{1}{\sqrt{6}} \left(
\Phi_{21}+\Phi_{32}+2\Phi_{31} \right), 
 \label{pi2-global}
\end{equation}
with $t_1<t_2<t_3$. If Eq.~(\ref{hyp-global}) was an exact identity,
i.e., if the fit of $C(t,t')$ by $g[\phi(t)-\phi(t')]$ had exactly zero
residuals, then $\sigma=0$ and the global two-time function would
verify the relation

\begin{equation}
g^{-1}[C(t_2,t_1)]+g^{-1}[C(t_3,t_2)]-g^{-1}[C(t_3,t_1)]=0.
\label{PhiL=0-global}
\end{equation}

A more extensive analysis of the fluctuations can be performed by
separating longitudinal and transverse components in Eqs.~(\ref{pi1})
and (\ref{pi2}). First, we recall that $\sigma$ is a purely
longitudinal quantity [see Eq.~(\ref{sigma})], therefore
$\sigma_{{\vec{r}}}{}^T=0$ and
$\Phi_{31,{\vec{r}}}{}^T=\Phi_{21,{\vec{r}}}{}^T +
\Phi_{32,{\vec{r}}}{}^T$. Now, we make the following two
assumptions. One is that the transverse and longitudinal fluctuations
are not correlated to each other. The other is that all slow
fluctuations are transverse, or in other words, that longitudinal
fluctuations are short range correlated in time, or at least they are
correlated over times that are shorter than the shortest time interval
between the configurations that are being considered. This leads to
the conditions

\begin{equation}
\left \langle\delta \Phi_{ab,{\vec{r}}}{}^L\delta \Phi_{cd,{\vec{r}}'}{}^T
\right\rangle=0 \quad\quad \forall \quad a,b,c,d, {\vec{r}}, {\vec{r}}' 
\label{def}
\end{equation}
and
\begin{equation}
\left \langle\delta \Phi_{ab,{\vec{r}}}{}^L\delta \Phi_{cd,{\vec{r}}'}{}^L
\right\rangle=0 \quad\quad \text{for} \quad a \not= c \quad \text{or} \quad
b \not= d, \quad\quad \forall \quad {\vec{r}}, {\vec{r}}'. 
\label{def1}
\end{equation}
Here $\delta x \equiv x-\langle x \rangle$. By using
Eq. (\ref{transver-eq}), it can be shown that the transverse
components of Eqs. (\ref{pi1}) and (\ref{pi2}) are given by

\begin{equation}
\pi_{1,{\vec{r}}}{}^T=\frac{1}{\sqrt{2}}(\Phi_{21,{\vec{r}}}{}^T-\Phi_{32,{\vec{r}}}{}^T) 
\end{equation}
and
\begin{equation}
\pi_{2,{\vec{r}}}{}^T=\frac{3}{\sqrt{6}}(\Phi_{21,{\vec{r}}}{}^T+\Phi_{32,{\vec{r}}}{}^T). 
\end{equation}

Regarding the longitudinal components, by using Eq. (\ref{def1}) it
can be shown that

\begin{equation}
\left\langle (\delta \sigma_{\vec{r}}{}^L)^2 \right\rangle= \frac{1}{3} \left[
  \left\langle (\delta \Phi_{21,{\vec{r}}}{}^L)^2 \right\rangle+\left\langle (\delta
  \Phi_{32,{\vec{r}}}{}^L)^2 \right\rangle+\left\langle (\delta
  \Phi_{31,{\vec{r}}}{}^L)^2 \right\rangle \right], 
\label{long-sigma}
\end{equation}

\begin{equation}
\left\langle (\delta \pi_{1,{\vec{r}}}{}^L)^2 \right\rangle= \frac{1}{2} \left[
  \left\langle (\delta \Phi_{21,{\vec{r}}}{}^L)^2 \right\rangle+\left\langle (\delta
  \Phi_{32,{\vec{r}}}{}^L)^2 \right\rangle \right],
\label{long-pi1}
\end{equation}
and
\begin{equation}
\left\langle (\delta \pi_{2,{\vec{r}}}{}^L)^2 \right\rangle= \frac{1}{6} \left[
  \left\langle (\delta \Phi_{21,{\vec{r}}}{}^L)^2 \right\rangle+\left\langle (\delta
  \Phi_{32,{\vec{r}}}{}^L)^2 \right\rangle+4\left\langle (\delta
  \Phi_{31,{\vec{r}}}{}^L)^2 \right\rangle \right]. 
\label{long-pi2}
\end{equation}

From the last three equations it can be easily seen that   
\begin{equation}
\left\langle (\delta \pi_{1,{\vec{r}}}{}^L)^2 +(\delta
\pi_{2,{\vec{r}}}{}^L)^2-2(\delta \sigma_{\vec{r}}{}^L)^2 \right\rangle=0. 
\label{long}
\end{equation}

By using Eq. (\ref{def}) we find that, all together, we can compute
the variance of the transverse fluctuations by combining the
fluctuations of Eqs. (\ref{sigma}), (\ref{pi1}), and (\ref{pi2}) in the
following way

\begin{equation}
\left\langle(\delta\pi_{1,{\vec{r}}}{}^T)^2+(\delta\pi_{2,{\vec{r}}}{}^T)^2\right\rangle=\left\langle(\delta\pi_{1,{\vec{r}}})^2+(\delta\pi_{2,{\vec{r}}})^2-2(\delta 
\sigma_{\vec{r}})^2\right\rangle.  
\label{var-tran}
\end{equation}

Further, we can estimate correlation volumes (in units of the coarse
graining volume $V_{cg}$) by using the formula
\begin{equation}
V_{\rm corr} \equiv \chi_{4,a}/ \left[ V_{cg} \left\langle(\delta
  a_{\vec{r}})^2\right\rangle \right],
\end{equation} 
where $\chi_{4,a} \equiv V \left\langle (\delta
\overline{a})^2\right\rangle$, $a_{\vec{r}}$ is a local coarse grained
variable, $\overline{a}\equiv \int \frac{d^dr}{L^d}a_{\vec{r}}$ is the
spatial average of $a_{\vec{r}}$, and $V =
L^d$ is the volume of the system.
% and $V_{cg}$ is the coarse graining volume. 
Therefore, by using these equations together with
Eq.~(\ref{var-tran}) we can estimate the correlation volume of
transverse and longitudinal fluctuations, respectively given by
\begin{eqnarray}
V^T=
\frac{\chi_{4,\pi_{1}{}^T}+\chi_{4,\pi_{2}{}^T}}{V_{cg}\left\langle(\delta
  \pi_{1,{\vec{r}}}{}^T)^2+(\delta
  \pi_{2,{\vec{r}}}{}^T)^2\right\rangle}\hspace{ 0.44 in} \nonumber \\ 
=\frac{V\left\langle(\delta \overline{\pi_{1}})^2+(\delta
  \overline{\pi_{2}})^2-2 (\delta
  \overline{\sigma})^2\right\rangle}{V_{cg}\left\langle(\delta
  \pi_{1,{\vec{r}}})^2+(\delta \pi_{2,{\vec{r}}})^2-2 (\delta
  \sigma_{{\vec{r}}})^2\right\rangle} 
\label{vol-trans}
\end{eqnarray}
and

\begin{equation}
V^L=\frac{\chi_{4,\sigma}}{V_{cg} \left\langle(\delta
  {\sigma_{\vec{r}}})^2\right\rangle}=\frac{V \left\langle(\delta
  \overline{\sigma})^2\right\rangle}{V_{cg} \left\langle(\delta
  \sigma_{\vec{r}})^2\right\rangle}. 
\label{vol-long}
\end{equation}

If the time reparametrization hypothesis is correct, we expect that
the variance as well as the correlation volume of the transverse
fluctuations will grow together with those corresponding to the
dynamical heterogeneities. We also expect that the variance and
correlation volume of the longitudinal fluctuations should be
insensitive to changes in the variance and correlation volume of the
dynamical heterogeneities.

\section{Testing the Hypothesis}\label{systems}

\subsection{Systems}
We performed classical molecular dynamics simulations
%~\cite{previous-reports} 
in systems of $N$ particles that were initially equilibrated at high
temperature $T_i \gg T_c$ (where $T_c$ is the mode coupling critical
temperature~\cite{Bengtzelius1984}), then instantaneously quenched to
a final temperature $T$ and allowed to evolve for times several orders
of magnitude longer than their typical vibrational
times~\cite{Castillo2007, Parsaeian2008, Parsaeian2009,
  Parsaeian2008a}. We determined $T_c$ by fitting
$\tau_{\alpha}=(T-T_{c})^{\gamma}$, where $\tau_{\alpha}$ is the
equilibrium $\alpha$-relaxation time defined by the conditions
$C(t_1+\tau_\alpha(t_1), t_1)= 1/e$ and $\tau_{\alpha} = \lim_{t_1 \to
  \infty} \tau_\alpha(t_1)$, as shown in Ref.~\cite{Parsaeian2009}. We
generated eight datasets by simulating four atomistic glass-forming
models~\cite{Parsaeian2008a}. Two of the systems are 80 : 20 mixtures of
A and B particles, interacting via either Lennard-Jones (LJ)
potentials~\cite{Kob-Andersen} or via purely repulsive
Weeks-Chandler-Andersen (WCA) truncation of the LJ potentials ~\cite{WCA}. The interactions
in the particle systems have the same length parameters,
$\sigma_{\alpha\beta}$ ($\alpha, \beta \in \{A,B\}$), and energy
parameters, $\epsilon_{\alpha\beta}$ as in the standard Kob-Andersen
mixture, namely~\cite{Kob-Andersen}: $\sigma_{AA}=1.0$,
$\sigma_{AB}=0.8$, $\sigma_{BB}=0.88$, $\epsilon_{AA}=1.0$,
$\epsilon_{AB}=1.5$, and $\epsilon_{BB}=0.5$. The other two systems are
models of short (ten-monomer) polymers, in which all particles interact
with each other via either LJ potentials or via WCA potentials, with
length parameter $\sigma_{AA}=1.0$ and energy parameter
$\epsilon_{AA}=1.0$. Additionally, in our polymer models,
nearest-neighbor monomers along a chain are connected by a FENE
anharmonic spring potential. For both particle and polymer systems,
the LJ potential is truncated at the cut-off distance $r_{\rm cutoff,
  \alpha \beta} = 2.5 \, \sigma_{\alpha\beta}$ and the WCA potential is
truncated at $r_{\rm cutoff, \alpha \beta} = 2^{1/6}\sigma_{\alpha
  \beta}$. We choose the unit of length as $\sigma_{AA}$, the unit of
energy as $\epsilon_{AA}$ and the unit of time as
$(\sigma^2_{AA}M/48\epsilon_{AA})^{1/2}$. For the particle systems,
the simulations were performed in an NVT ensemble, with the
temperature being fixed by the rescaling method and being checked every 50 timesteps. For the polymer
systems, the simulations were performed in an NPT ensemble, with both
the pressure and the temperature being controlled by the Nose-Hoover
method. The details of the simulations are summarized in
Table~\ref{tabla}.
\begin{table}[h]% add [H] placement to break table across pages
\begin{ruledtabular}
\centering
\begin{tabular}{ c  c c c r c l c}
\multicolumn{8}{c}{\large{\textbf{Systems}}}\\[0.5ex]
 & & & & &\\ 
%\hline
\multicolumn{1}{b{1.5cm}}{\begin{center}Label\end{center}} &
 \multicolumn{1}{b{1.5cm}}{\begin{center}$N$\end{center}} &
   \multicolumn{1}{b{1.5cm}}{\begin{center}$m$\end{center}} &
   \multicolumn{1}{b{1.5cm}}{\begin{center}Potential\end{center}} &
   \multicolumn{1}{b{.6cm}}{\begin{center}$T$\end{center}} &
   \multicolumn{1}{b{.5cm}}{\begin{center}$T_c$\end{center}} &
   \multicolumn{1}{b{.5cm}}{\begin{center}$T/T_c$\end{center}} &
   \multicolumn{1}{b{1.5cm}}{\begin{center}Runs\end{center}} \\ 
   %Label & $N$ & $m$ &Potential &$T/T_c$ &Runs\\ [1ex]
\hline
& & & &  & &\\[-1ex]
A   &  8000  &   10 &  LJ & 0.6 & 0.833&$\approx 0.7$ & 100 \\ [0.5ex]%0.6
B   & 8000  &    10 &WCA &0.4&0.503& $\approx 0.8$ &  800\\ [0.5ex]%0.4
C   & 8000  &     1& LJ &  0.4&0.435&$\approx 0.9$& 250\\ [0.5ex]%0.4
D   & 1000  &     1& WCA & 0.236 &0.263&$\approx 0.9$ & 5000\\ [0.5ex]%0.2367
E   & 1000  &     1 & WCA &  0.29 &0.263&$\approx 1.1$  & 9000\\ [0.5ex]%0.29
F   & 1000  &     1& WCA - Eq &0.29&0.263& $\approx 1.1$  & 9000\\ [0.5ex]%0.29
G   & 1000  &     1 &WCA &0.4&0.263&$\approx 1.5$  & 4999\\ [0.5ex]%0.4
H   & 1000  &     1& WCA - Eq &0.4&0.263&$\approx1.5$ & 4999\\ [1ex]%.4
\end{tabular}
\end{ruledtabular}
\caption{ Details of the numerical simulations analyzed in this
  work. We considered systems of $N$ particles, with $m$ particles per
  molecule, interacting via either Lennard-Jones (LJ) potentials or
  via purely repulsive Weeks-Chandler-Andersen (WCA) potentials, at
  final temperature $T$. Each temperature is also described by its
  ratio with respect to the empirically determined mode coupling
  critical temperature $T_c$~\cite{Bengtzelius1984} for the same
  system. The last column lists the number of independent
  runs. Datasets F and H correspond to systems in equilibrium, all
  others to systems in the aging regime.}{\label{tabla}}
\end{table}

\subsection{Results}

\begin{figure}[h!]
 \begin{center}
 \includegraphics[scale=.34]{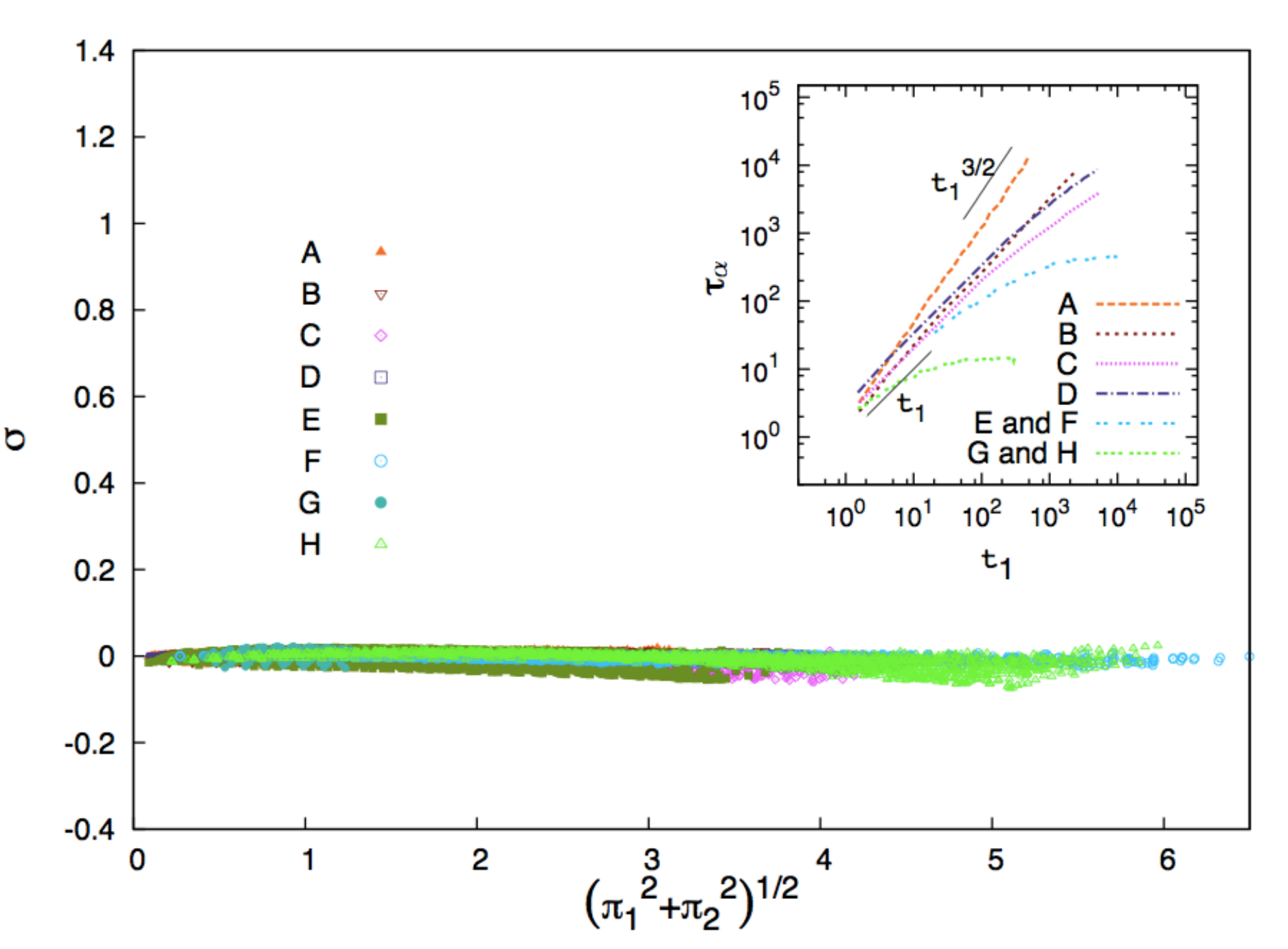}%
 \caption{ (Color online) Global $(\sigma, \sqrt{\pi_1^2+\pi_2^2})$
   pairs for all data sets and all possible times $t_1 < t_2 <
   t_3$. {Inset:} Relaxation time $\tau_{\alpha}(t_1)$ as a function of the
   waiting time $t_1$ for all the models and temperatures
   considered. $\tau_{\alpha}(t_1)$ is defined by the condition
   $C(t_1+\tau_{\alpha}(t_1), t_1)= 1/e$. }
 \label{global}
 \end{center}
 \end{figure}

We begin by identifying the functions $g(x)$ and $\phi(t)$
[Eq. (\ref{hyp-global})] that best describe the global correlations
$C(t,t')$ computed from our data sets by using
Eq. (\ref{global-correlation}). We find that, for all data sets, the
two-time correlation $C(t,t')$ can be well fitted by using the form
$g(x)=q_{EA} \exp \left[ -(x/ \theta_0)^{\beta} \right]$. However, as
shown in the inset of Fig.~\ref{global}, the relaxation for different
systems presents different behaviors, which leads to different forms
of $\phi(t)$. The best fits of $C(t,t')$ that we obtained correspond
to the following forms: for aging polymer systems $\phi(t)=\ln
^{\alpha} (t/t_0)$, for aging particle systems
$\phi(t)=(t/t_0)^{\alpha}$, and for equilibrium $\phi(t)= t/t_0$. We
can verify our proposed Eq.~(\ref{hyp-global}) for the different cases
by using the form of the obtained functions $g(x)$ and $\phi(t)$. For
equilibrium we trivially recovered, as expected, the case of TTI. In
the aging cases we can verify that Eq.~(\ref{hyp-global}) can be
rewritten in the form $C(t,t') = f\left[ h(t)/h(t') \right]$
\cite{Avila}, which is found in many aging
systems~\cite{Bouchaud}. Once the fitting procedure is performed for a
given dataset, we can use the known values of the parameters $q_{EA}$,
$\beta$ and $\theta_0$ to compute
\begin{equation}
\Phi_{ab}=g^{-1}[C(t_a,t_b)]=\theta_0\{-\ln[q_{EA}^{-1}C(t_a,t_b)]\}^{1/\beta}
\end{equation}
and, by using Eqs. (\ref{sigma-global})--(\ref{pi2-global}), to compute
$\sigma$, $\pi_1$, and $\pi_2$. Figure~\ref{global} shows the results of
plotting the global values of $\sigma$ against the global values of
$\sqrt{\pi_1^2+\pi_2^2}$ for all times, $t_1 < t_2 < t_3$, and all
systems. As discussed before, since the fits are not perfect, the
results do not fall exactly on the line $\sigma=0$, but the collapse
and the fits are good enough to allow us to test the hypothesis.

\begin{figure}[h!]
 \begin{center}
 \includegraphics[scale=.51]{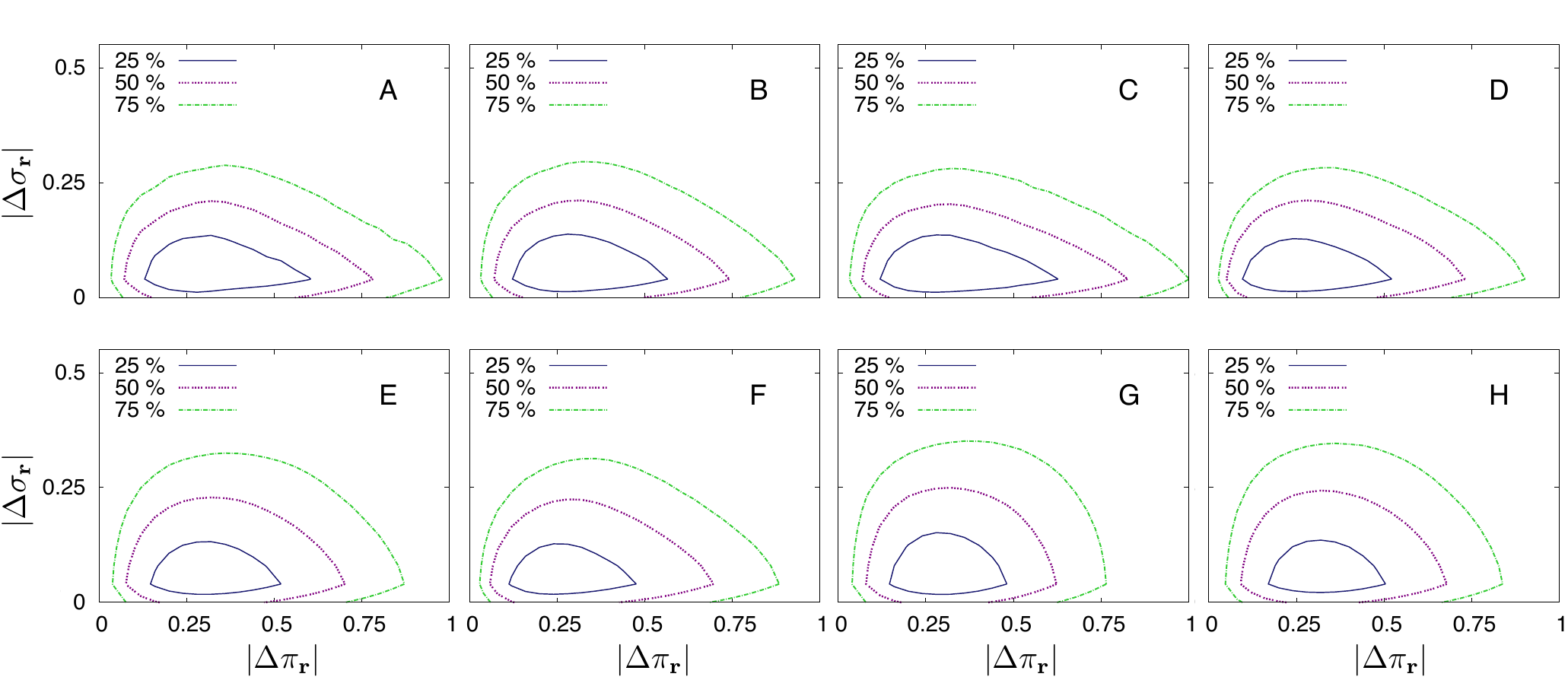}%
 \caption{ (Color online) 2D contours of constant joint probability
   density $\rho(|\Delta \sigma_{\vec{r}}|, |\Delta \pi_{\vec{r}}|)$,
   computed using coarse graining boxes containing 125 particles on
   average. By subtracting the global quantities to the local
   quantities we avoid trivial effects due to differences in the
   global values. Each panel from A to H contains results from the
   corresponding dataset, for $C(t_3,t_1)\approx0.23$, with the times
   chosen as late as possible within each dataset. Each set of three
   concentric contours is chosen so that they enclose $25\%, 50\%$, and
   $75\%$ of the total probability.}
 \label{local-contours}
 \end{center}
 \end{figure}

\begin{figure}[h]
  \centering
  %\subfloat[]{\label{fig:contours-C23}\includegraphics[scale=.65]{WCA-C13-023-4.pdf}} 
  %\subfloat[]{\label{fig:contours-C33}\includegraphics[scale=.65]{WCA-C13-033-4.pdf}}\\
  %\subfloat[]{\label{fig:contours-C44}\includegraphics[scale=.73]{WCA-C13-044-4.pdf}}
  \includegraphics[scale=0.505]{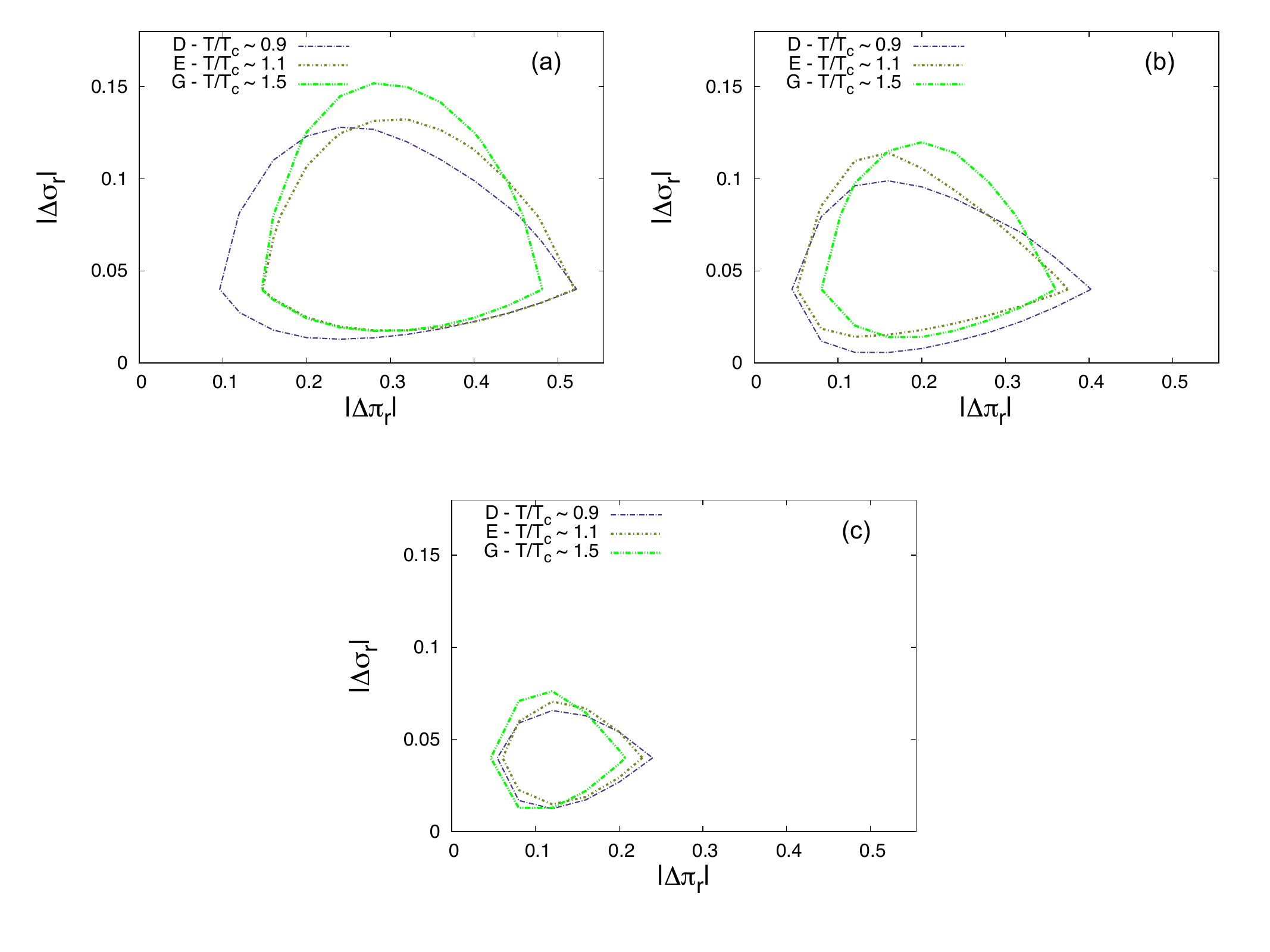}
  \caption{(Color online) Comparison of the contours enclosing $25\%$
    of the probability for the aging WCA systems at three different
    temperatures (D, E, and G). The three different panels correspond
    to different values of $C(t_3,t_1)$: (a) $C(t_3,t_1)\approx0.23$,
    which is the same value as in Fig.~\ref{local-contours}, (b)
    $C(t_3,t_1)\approx0.33$, and (c) $C(t_3,t_1)\approx0.44$.}
  \label{fig:comparison}
\end{figure}

As mentioned in the previous section, if Eq.~(\ref{hyp0}) is
satisfied, we expect the local quantities to satisfy
$\sigma_{\vec{r}}=0$. In Fig.~\ref{local-contours} we show the results
of plotting for all systems the 2D projection of the joint probability
density of the coarse-grained local correlations, $\rho(|\Delta
\sigma_{\vec{r}}|, |\Delta \pi_{\vec{r}}|)$, with $|\Delta
\sigma_{\vec{r}}| \equiv |\sigma_{\vec{r}}-\sigma|$ and $|\Delta
\pi_{\vec{r}}| \equiv
\sqrt{(\pi_{1,{\vec{r}}}-\pi_1)^2+(\pi_{2,{\vec{r}}}-\pi_2)^2}$. The
values of the global quantities are subtracted from the local
quantities to avoid trivial effects due to differences in the global
values. By doing this we are able to better compare the contours
independently of the choice of $C(t_2,t_1)$ and $C(t_3,t_2)$. We do,
however, keep the value of $C(t_3,t_1)$ approximately the same for all
systems, in this case $C(t_3,t_1)\approx 0.23$. The three contours
shown for each dataset enclose respectively $25 \%$, $50 \%$, and
$75\%$ of the total probability. We coarse grain over moderately large
regions, containing on average 125 particles, in order to detect
collective fluctuations and, since time reparametrization symmetry is
a long time asymptotic effect, we choose the times as late as
possible.  As the time reparametrization hypothesis predicts, for
data sets with $T<T_c$ (A--D), the purely longitudinal fluctuations
${\Delta \sigma_{\vec{r}}}$ are clearly smaller than the $\Delta
\vec{\pi}_{\vec{r}}$ fluctuations, which contain both transverse and
longitudinal contributions. This behavior is more noticeable for the
$25\%$ contour, which encloses the most likely fluctuations, than for
the $50\%$ and $75\%$ contours. For moderately higher temperature,
$T/T_c \approx1.1$ (data sets E and F), the anisotropy $|\Delta \sigma
_{\vec{r}}|< |\Delta \vec{\pi}_{\vec{r}}|$ is still present, but less
pronounced. In the case of the highest temperature, $T/T_c \approx
1.5$ (data sets G and H), the anisotropy is either very slight, or
absent. In the case of the systems that are equilibrated, F and H, we
find that the shapes of their contours are similar but slightly more
anisotropic then the ones obtained for the same temperature in the
aging regime, E and G, respectively.  The effect of temperature in the
anisotropy of the contours can be observed in more detail in
Fig.~\ref{fig:comparison}(a) where the $25 \%$ contour of the
probability density for the systems of particles with WCA interactions
is shown for three temperatures (data sets D, E, G). This can be
directly connected to the fact that, as the temperature is increased,
the separation of timescales is less pronounced, the finite time
corrections to the time reparametrization symmetry become larger, and
the effects of local time variable fluctuations become weaker. The
same trends can be observed in Figs.~\ref{fig:comparison}(b) and
\ref{fig:comparison}(c) for the same datasets as in
Fig.~\ref{fig:comparison}(a) but for different values of the global
correlation $C(t_3,t_1)$.

\begin{figure}[h]
 \begin{center}
 \includegraphics[scale=.87]{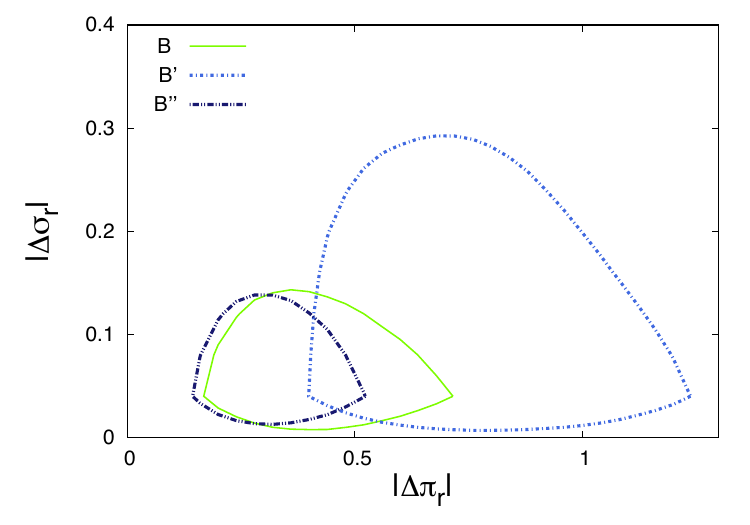}%
 \caption{(Color online) $25\%$ probability contours for dataset B
   with $C(t_3,t_1)\approx0.23$. Contour B corresponds to the $25 \%$
   of the probability density shown in
   Fig.~\ref{local-contours}. Contour B' corresponds to the same times
   as contour B, but with a much smaller coarse graining size. Contour
   label B'' corresponds to the same coarse graining size as B, but
   with much shorter times.}
 \label{diff}
 \end{center}
 \end{figure}

We can further analyze the effects of choosing different conditions
from the ones chosen in Fig.~\ref{local-contours}, for instance, by
comparing the results shown in Fig.~\ref{local-contours} with results
obtained for smaller coarse graining regions or for shorter times in
the aging regime. Exactly this kind of comparison is shown in
Fig.~\ref{diff}, where the $25 \%$ probability contours for dataset B
are shown for three conditions. The contour labeled B is the one shown
already in Fig.~\ref{local-contours}. The contour labeled B'
corresponds to the same time, but with coarse graining regions
containing on average 23 particles instead of 125. This leads to less
averaging and stronger fluctuations, but also, since fluctuations
correlated over shorter distances are no longer preferentially
suppressed, the shape of the contour is no longer dominated by
collective modes, and thus contour B' extends more in the direction of
$|\Delta \sigma_{\vec{r}}|$ than contour B. The contour labeled B''
corresponds to the same coarse graining size of contour B, but with
much shorter times. This leads to stronger finite time effects,
analogous to the ones found at slightly higher temperatures, and as
expected the contour is less anisotropic, and indeed, it resembles the
contours corresponding to $T /T_c \approx 1.1$.

\begin{figure}[h!]
 \includegraphics[scale=0.59]{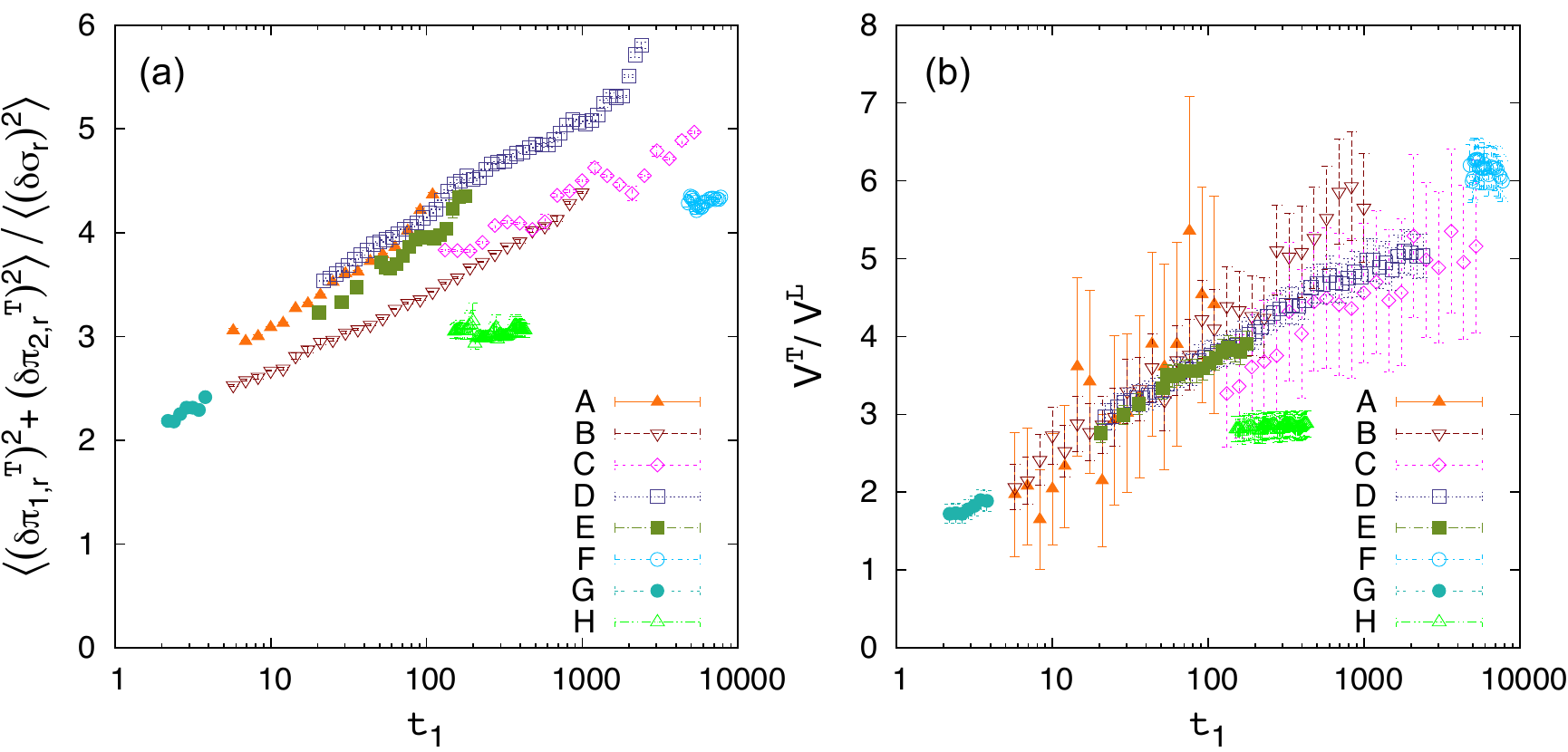}
  \caption{(Color online) (a) Ratio between the variances of
    transverse and longitudinal fluctuations as a function of the
    initial time $t_1$, for $C(t_3, t_1) \approx 0.23$. (b) Ratio between the correlation volumes of
    transverse and longitudinal fluctuations as a function of
    $t_1$, for the value $C(t_3, t_1) \approx 0.23$.}%The color coding identifies the datasets and is the same
           %as in Fig.~(\ref{global}).}.
  \label{fig:evolution-tw}
\end{figure}

\begin{figure}[h!]
  \includegraphics[scale=0.6]{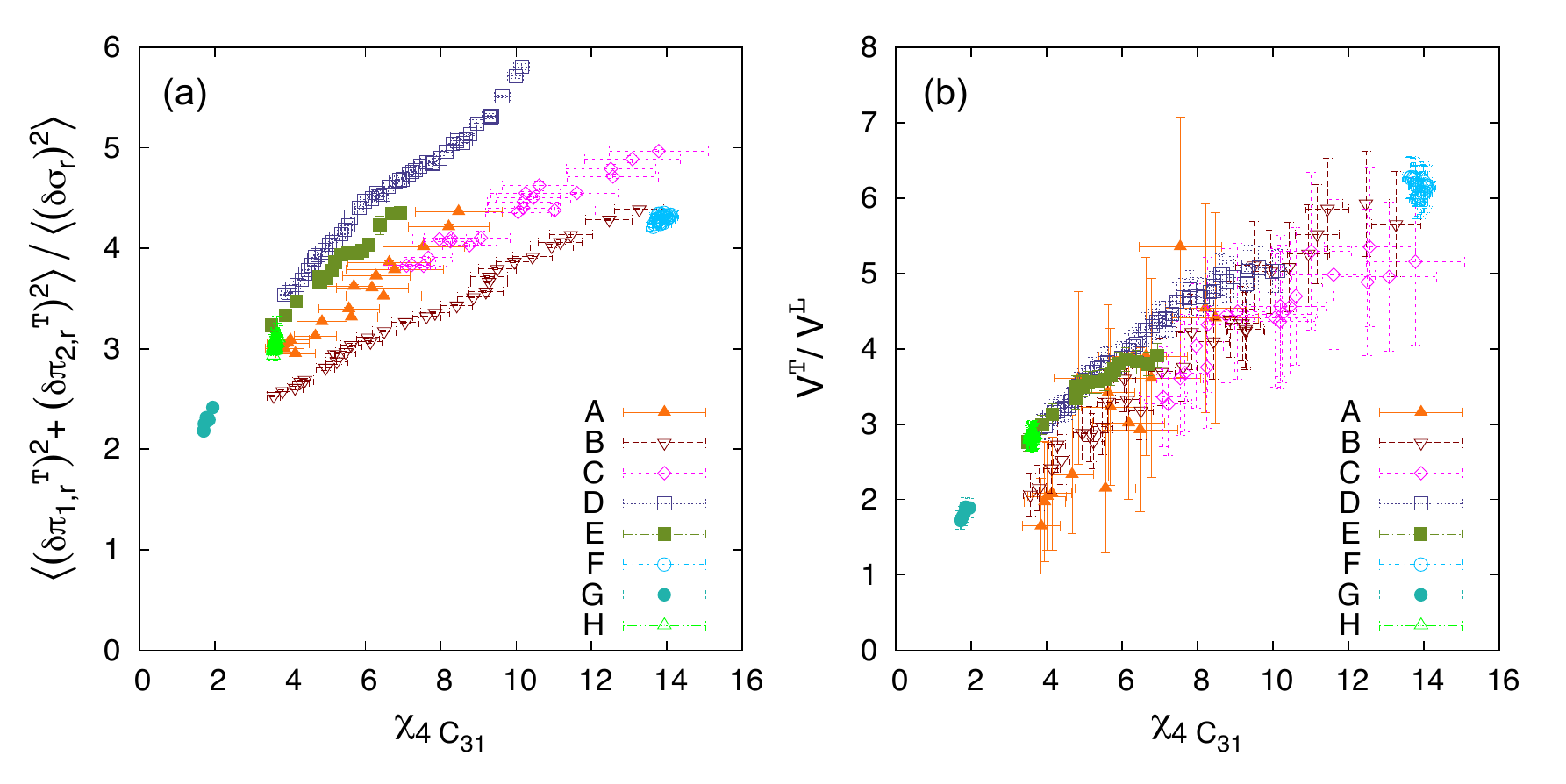}
  \caption{(Color online) (a) Ratio between the variances of
    transverse and longitudinal fluctuations as a function of the
    strength of the dynamical heterogeneity, measured by $\chi_{4,
      C_{31}} \equiv \chi_4(t_3,t_1)$, for the value $C(t_3, t_1) \approx 0.23$. (b) Ratio between the
    correlation volumes of transverse and longitudinal fluctuations as
    a function of $\chi_{4,C_{31}}$, for $C(t_3, t_1) \approx 0.23$.} 
  % The color coding is the same as in Fig.~(\ref{global}).}.

  \label{fig:evolution}
\end{figure}

 We now move to a more quantitative analysis of the {\em strength} and
 {\em spatial correlations\/} of the fluctuations by making use of the
 results derived in Sec.~\ref{TRI}.  In Fig.~\ref{fig:evolution-tw}(a)
 we show the ratio between the variances of local transverse
 fluctuations and longitudinal fluctuations,
 $\left\langle(\delta\pi_{1,{\vec{r}}}{}^T)^2+(\delta\pi_{2,{\vec{r}}}{}^T)^2\right\rangle/
 \left\langle(\delta \sigma_{\vec{r}})^2\right\rangle$ (``variance anisotropy
 ratio") [see Eq. (\ref{var-tran})], as a function of the initial time
 $t_1$. 
Similarly, in Fig.~\ref{fig:evolution-tw}(b) we plot the ratio
 between the correlation volumes of transverse and longitudinal
 fluctuations, $V^T/V^L$ (``correlation volume anisotropy ratio") [see
 Eqs. (\ref{vol-trans}) and (\ref{vol-long})], also as a function of
 $t_1$. We find that for aging systems both ratios grow as $t_1$
 increases, as one could expect from the fact that at later times the
 reparametrizations symmetry breaking terms in the action should
 become progressively weaker~\cite{Chamon2002}. For the equilibrium
 datasets, the dynamics is time translation invariant (TTI), and we
 observe, as expected, that both anisotropies are independent of
 $t_1$.  In Fig.~\ref{fig:evolution}, we show the same two ratios as
 functions of the strength of the dynamical heterogeneities, measured
 by the dynamical susceptibility $\chi_4(t_3,t_1) \equiv \chi_{4,
   C_{31}} = N \left[ \left\langle \overline{C_{31}}^2 \right\rangle -\left\langle
   \overline{C_{31} }\right\rangle^2 \right]$. We observe that
 both anisotropy ratios grow when $\chi_{4, C_{31}}$
 increases, i.e., as the dynamical heterogeneity becomes more
 pronounced. Although the same qualitative behavior is observed for
 all datasets, the curves are different for different systems and
 temperatures.

\begin{figure}[h!]
  \begin{center}
    \includegraphics[scale=1.05]{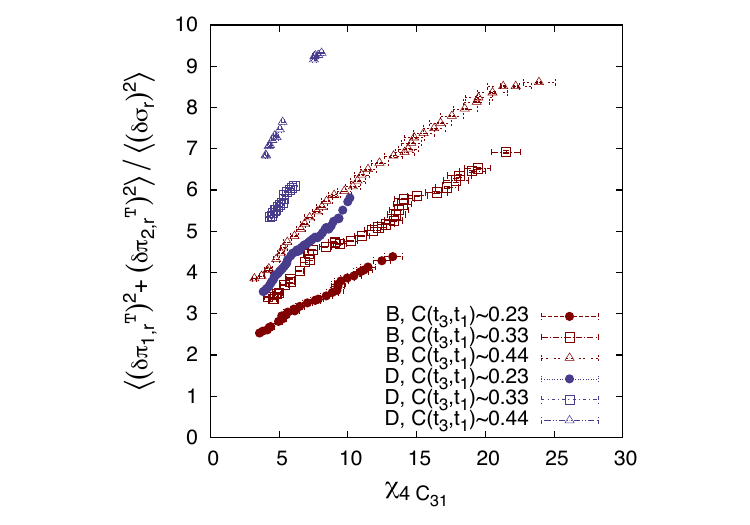}
  \end{center}
  \caption{(Color online) Ratio between the variances of transverse
    and longitudinal fluctuations as a function of $\chi_{4, 
      C_{31}}$, for systems B and D for the values
    $C(t_3,t_1)\approx0.23$, $0.33$, and $0.44$. }%The color coding is
                                %the same as in Fig.~(\ref{global}).} 
    \label{fig:diff-C13}
\end{figure}

The results presented in Fig.~\ref{fig:evolution} correspond to the
value $C(t_3,t_1) \approx 0.23$, but similar results can be obtained for
different values of $C(t_3,t_1)$, as one can expect from
Fig.~\ref{fig:comparison}. Results for three different values of
$C(t_3,t_1)$ are shown in Fig.~\ref{fig:diff-C13}, where the variance
anisotropy ratio for datasets B and D is plotted as a function of
$\chi_{4, C_{31}}$ for the values $C(t_3,t_1)\approx$ $0.23$,
$0.33$, and $0.44$. This figure shows that the variance anisotropy
ratio grows with the strength of the dynamical heterogeneity, for all
three fixed values of $C(t_3,t_1)$.

\begin{figure}[h]
  \begin{center}
    \includegraphics[scale=0.45]{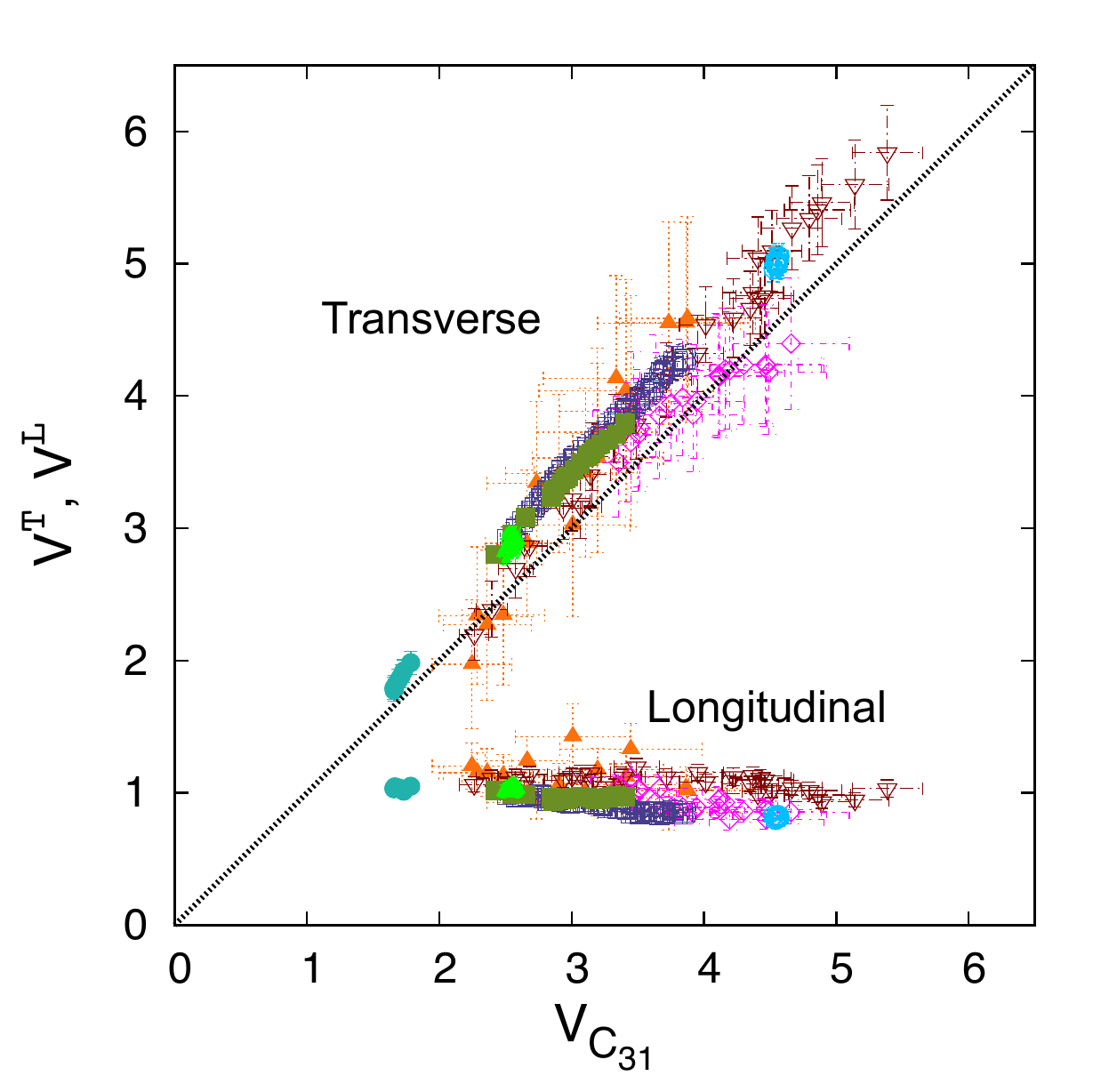}
  \end{center}
  \caption{(Color online) Correlation volumes for transverse and
    longitudinal fluctuations, plotted against $V_{C_{31}}$, the
    correlation volume for the fluctuations of $C_{31,{\vec{r}}}$, calculated for the value $C(t_3, t_1) \approx 0.23$.}

  \label{fig:corr-vol}
\end{figure}

According to the hypothesis, dynamical heterogeneity originates in the
Goldstone modes associated to fluctuations in the time
reparametrization [see Eq.~(\ref{hyp})]. Therefore, the hypothesis
implies that the correlation volume of the dynamical heterogeneity
should be similar to the correlation volumes of the transverse
components of the variables $\pi_1$ and $\pi_2$, and the longitudinal
variable $\sigma$ should be less correlated in space.  Our results,
shown in Fig.~\ref{fig:corr-vol}, show that this is indeed the
case. The correlation volume corresponding to the transverse
fluctuations, $V^T$, closely tracks the one corresponding to the
dynamical heterogeneities, $V_{C_{31}}$, and they both grow together
as the temperature is reduced or the timescale is increased. By
contrast, for the longitudinal fluctuations $\sigma$ we find that
their correlation volume $V^L$ is small and essentially constant; it does
not correlate with the correlation volume of the dynamical
heterogeneity, nor with the temperature or the time scale. In fact,
despite the large error bars and the presence of some outlier points,
the figure shows a partial data collapse between different systems,
both for the case of transverse and for the case of longitudinal
correlation volumes. In the case of longitudinal fluctuations, this
may be a trivial effect due to the correlation volumes being smaller
than the volume of the coarse graining regions used to define the
variables. In the case of the transverse fluctuations, the correlation
volumes go well beyond the volume of the coarse graining regions, and
the partial collapse in the results might be evidence of some sort of
universality, but more work will be needed in order to decide this
question one way or another.

\section{Conclusions}\label{conclusions}

In this paper we tested the hypothesis that dynamical heterogeneity
arises from Goldstone modes related to a broken continuous symmetry
under time reparametrizations. In other words, we tested whether
dynamical heterogeneity is associated with the presence of spatially
correlated fluctuations in the time variables. To verify this, we
studied probability distributions that allowed us to distinguish
between time reparametrization fluctuations (transverse fluctuations)
and other fluctuations (longitudinal fluctuations). We also tested for
possible correlations of both the strength and the correlation volume
of the fluctuations in the time variable with the dynamical
susceptibility $\chi_4$, which is normally used to probe dynamical
heterogeneity. Altogether, we found that at the lowest temperatures,
for the longest timescales and for the largest coarse graining
lengths, the transverse fluctuations became stronger than the
longitudinal fluctuations, which is consistent with the hypothesis. We
also found that the correlation volumes of the time reparametrization
fluctuations were proportional to the correlation volume of the
dynamical heterogeneity, while the correlation volumes of the
longitudinal fluctuations were small and independent of the
correlation volumes of the dynamical heterogeneity. These observations
apply to all the systems examined, regardless of the details of the
interaction (purely repulsive in the case of WCA vs.\ repulsive and
attractive in the case of LJ), the kinds of objects (binary systems of
particles vs.\ systems of short polymers), or the ensembles used in
the simulations (NVT vs.\ NPT). All of this was despite the fact that,
to simplify the analysis, we imposed some extra conditions on the form
of the correlations, which may have made the agreement with the
hypothesis appear less good than it would have been otherwise.

With regards to universality, the evidence we found was mixed. On the
one hand, there were clear differences in the details of the results
for different systems, for example for the anisotropy ratios. On the
other hand, all the trends we observed were the same across systems,
and the results for the correlation volumes did show some hints of
universality, although the relatively large noise in this measurement
did not allow for definite conclusions to be drawn. In any case, the
commonality in the results is strong enough to suggest that other
systems may display similar qualitative behaviors. Thus, we expect
that it would be very instructive to apply the same kind of test to
data from other slowly relaxing systems, such as particle tracking
data from glassy colloids~\cite{Kegel2000, Weeks2000, Weeks2002,
  Courtland2003} and from granular systems close to
jamming~\cite{Keys2007, Abate2007}.

Finally, considering the success of the tests presented here, it
becomes natural to ask if it is possible to extract from the data the
actual fluctuating reparametrization $\phi_{\vec{r}}(t)$, and to study
its properties directly. In fact, Ref.~\cite{Mavimbela} shows that
some progress can be made in that direction.

\section{Acknowledgments}\label{acknowledgments}
H.E.C. thanks E.~Flenner, M.~Kennett, G.~Szamel, and E.~Weeks for
discussions and C.~Chamon and L.~Cugliandolo for suggestions and for
many discussions. This work was supported in part by the DOE under grant no.
DE-FG02-06ER46300, by NSF under grants no. PHY99-07949 and PHY05-51164,
and by Ohio University. Karina Avila acknowledges the Condensed Matter and Surface Sciences (CMSS) program for a studentship. Numerical simulations were carried out at the
Ohio Supercomputing Center.  H.E.C. acknowledges the hospitality of
the Aspen Center for Physics and the Kavli Institute for Theoretical
Physics, where parts of this work were performed.


\begin{thebibliography}{10}
\bibitem{Notation} 
  Our notation differs from the one used in~\cite{Avila}. Here, we use
  $C_{\vec{r}}(t,t') \approx g[\phi_{\vec{r}}(t)-\phi_{\vec{r}}(t')]$
  instead of
  $C_{\vec{r}}(t,t') \approx f[h_{\vec{r}}(t)/h_{\vec{r}}(t')]$. This
  corresponds to the changes of variables $g(x)=f[\exp(x)]$ and
  $\phi_{\vec{r}}=\ln(h_{\vec{r}})$. Also, in this work, our
  definition of $\Phi_{ab}$ corresponds to the logarithm of the
  definition of $\Phi_{ab}$ used in~\cite{Avila}.
\end{thebibliography}
\end{document}